\documentclass{article}
\usepackage{authblk}
\usepackage{graphicx}
\title{Implementation of X-rays production in LegPy }
\author{Víctor Moya}
\author{Jaime Rosado}
\author{Fernando Arqueros}
\affil{\small IPARCOS-UCM, Instituto de Física de Partículas y del Cosmos and EMFTEL Department. Universidad Complutense de Madrid. Madrid 28040. Spain}
\date{}
\begin{document}
\maketitle
\begin{abstract}
The LegPy package has been upgraded by including the production of X-ray fluorescence based on a simple model. Several tests have been done to check the validity of our approximations. This simple model has been sufficient to fix severe deviations of LegPy with respect to the well-established code PENELOPE in those particular cases for which fluorescence emission has a relevant impact.
\end{abstract}

\section{Introduction}
LegPy (Low energy gamma-ray simulation with Python) is a simple Monte Carlo (MC) algorithm for the simulation of the passage of low-energy gamma rays and electrons through any material medium. As described in detail in \cite{legpy}, the algorithm includes several approximations that accelerate the simulation while maintaining reasonably accurate results. Notably, pair production and Bremsstrahlung are ignored, which limits the applicability of LegPy to low energies. In spite of that, deviations in the deposited energy with respect to well established MC codes are smaller than 10\% in light media up to about 5 MeV. 

Ignoring X-ray fluorescence (atomic de-excitation after photoelectric absorption) was a further simplification in the original code \cite{legpy}. However, as is well known \cite{subbaiah}, this process has a relevant impact in heavy atoms if the incident photon energy is slightly higher than that of the K edge of the atomic photoelectric cross section. As shown in \cite{legpy}, dose buildup factors obtained with PENELOPE \cite{penelope} for lead at 100 keV are significantly larger than that obtained with LegPy.
Therefore, this simplification severely limited the applicability of LegPy in heavy atoms even at low energies.

In this report we describe the implementation of the X-ray production that we have carried out in LegPy (section 1). This implementation is currently included in the latest version of LegPy in the GitHub repository \cite{github}. Following its original purpose a number of approximations have been applied for the simulation of the production of X-rays in such a way that LegPy keeps providing an easy-to-use framework useful for applications that do not require the level of detail of available well-established MC programs. As will be shown in section 2, several tests have been done to check the validity of our approximations.

\section{A simple model for X-ray fluorescence}
\label{model}

Our model follows the procedure shown in the well-known book of F.H. Attix \cite{Attix}, although we only consider fluorescence from the atomic K-shell. Let us assume that a photon undergoes photoelectric absorption by an atom. If the photon energy $E$ is larger than the K-shell binding energy $E_{\rm k}$, a hole can be produced in this shell with a probability $P_{\rm k}$. The hole can be filled by an electron of an outer shell giving rise to fluorescence emission with a probability $Y_{\rm k}$, called fluorescence yield. In our algorithm, these parameters are taken from
\cite{xrays} ($E_{\rm k}$), \cite{Hubbell}, \cite{McMaster} ($P_{\rm k}$) and \cite{Lederer} ($Y_{\rm k}$) as functions of the atomic number $Z$.

The approximations used in our model ignore specific features of the atomic structure, allowing a simple implementation of the fluorescence emission in LegPy. In addition, only atomic de-excitations via the transitions (L$\rightarrow$K) and (M$\rightarrow$K), that is the $K_{\alpha}$ and $K_{\beta}$ lines, are considered since the probability of
transitions from outer shells is negligible. The X-ray energies $E_{K\alpha}$ and $E_{K\beta}$, as well as the relative intensities $I_{K\alpha}$ and $I_{K\beta}$ for each atom are taken from \cite{xrays}.

In case of photoelectric absorption in a medium containing different atoms (molecules or mixtures), the relative probability associated to each individual atom can be assumed to be proportional to $Z^m$ with $m$ in the range $3 - 4$ \cite{JC}. We use this approximation with $m = 3.5$ to randomly select the atom with which the photon interacts. Then, if $E>E_{\rm k}$ for this atom, it is decided whether an X-ray is emitted or not with probability $P_{\rm k} \cdot Y_{\rm k}$. If not emitted, a photoelectron with energy equal to $E$ is ejected\footnote{This is also an approximation since this energy can be shared with Auger electrons as well.} and either transported or forced to be absorbed at the interaction point, depending on the user choice. If an X-ray is emitted, it is decided whether it is a $K_{\alpha}$ or a $K_{\beta}$ photon and the remaining energy either $E - E_{K\alpha}$ or $E - E_{K\beta}$, respectively, is given to the photoelectron.

\section{Validation of the model}
\label{valid}
In order to validate the approximations described in the previous section, a set of simulations have been carried out to compare the results of the upgraded version of LegPy with those of the well established PENELOPE  code \cite{penelope} for a number of critical cases where X-ray fluorescence is expected to have a relevant impact. In section \ref{dose}, we show results of dose in depth for isotropic sources in lead and bone. In section \ref{2m}, the effect of X-ray fluorescence on the dose around a water-lead interface is studied. Lastly, in section \ref{scint} results on spectra of energy absorbed by two popular scintillators (NaI and BGO) are presented. 

\subsection{Isotropic source in a sphere}
\label{dose}
In these simulations an isotropic source is located at the center of a sphere of radius equal to five times the mean free path (mfp). The deposited dose is obtained as a function of the radial distance, $D(R)$. In the first place, the simulation has been carried out with LegPy using both options, with and without X-ray fluorescence. Then the same case is simulated with PENELOPE that includes, by default, X-ray fluorescence.

\begin{figure}[h]
\centering
\includegraphics[width=\linewidth]{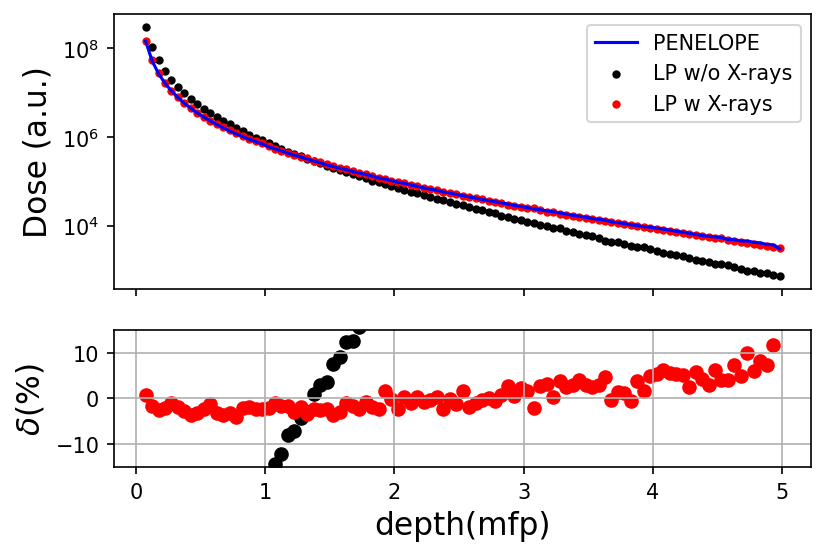}
\caption{Upper plot: Dose in depth for an isotropic source of 100 keV photons on a lead sphere obtained by LegPy compared with the predictions of PENELOPE. Lower plot: Percentage deviations of LegPy with (red points) and without (black points) X-ray fluorescence with respect to PENELOPE. See text for details.}
\label{fig:pb_100keV}
\end{figure}

The effect of X-rays might be particularly relevant in heavy media. For instance, the probability of X-ray emission after photoelectric absorption in lead is very high (75.8\%).
Since the cross section of this process strongly decreases with photon energy, it is expected a significant production of X-ray fluorescence when a beam of photons with energies slightly over the K-shell binding energy (88.0 keV) traverses lead. We have carried out the simulation for an isotropic source of 100 keV. The dose in depth $D(R)$ results are presented in figure \ref{fig:pb_100keV}. As pointed out in \cite{legpy}, very large deviations (by a factor of about 5 at 5 mfp) with respect to the predictions of PENELOPE are found when neglecting the fluorescence emission. On the other hand, it is shown that our implementation of the X-ray emission leads to results in good agreement with those of PENELOPE with deviations smaller than 10\% for depths up to 5 mfp.

As mentioned above, our implementation only accounts for the emission of $K_{\alpha}$ and $K_{\beta}$ lines, neglecting X-ray fluorescence  from photoelectric absorption producing holes in upper shells. To evaluate the impact of this approximation, we have compared our $D(R)$ results for 20 keV photons in lead (L-shell binding energy of 15.9 keV) with those of PENELOPE. The effect is not as strong as that of the K-shell fluorescence but still deviations of about 30\% are observed at 5 mfp.

\begin{figure}[h]
\centering
\includegraphics[width=\linewidth]{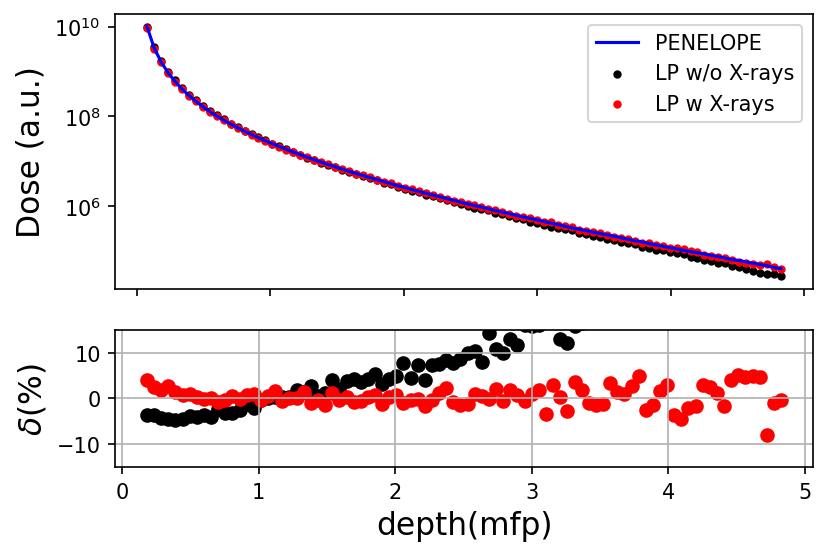}
\caption{Upper plot: Dose in depth for an isotropic source of 4.71 keV photons on a bone sphere obtained by LegPy compared with the predictions of PENELOPE. Lower plot: Percentage deviations of LegPy with (red points) and without (black points) X-ray fluorescence with respect to PENELOPE. See text for details.
}
\label{fig:bo_4.71}
\end{figure}

The effect of X-ray fluorescence might be also relevant for lighter elements. We have carried out simulations for bone, which is a medium of interest in medical physics. In particular, we used the so-called Cortical Bone (ICRU-44) with density and composition taken from \cite{icru-44}. In figure \ref{fig:bo_4.71}, results for the dose in depth for an isotropic source of 4.07 keV just above the K-shell energy of Ca ($E_{\rm k}$ = 4.04 keV) are shown. The effect of the K-shell fluorescence of Ca can be appreciated. Significant deviations of LegPy with respect to PENELOPE (up to 30\% at 5 mfp) are observed when ignoring fluorescence. These deviations are significantly reduced ($<$ 5\%) when including our simple X-ray fluorescence model.     

\subsection{Two media}
\label{2m}

\begin{figure}[h]
\centering
\includegraphics[width=\linewidth]{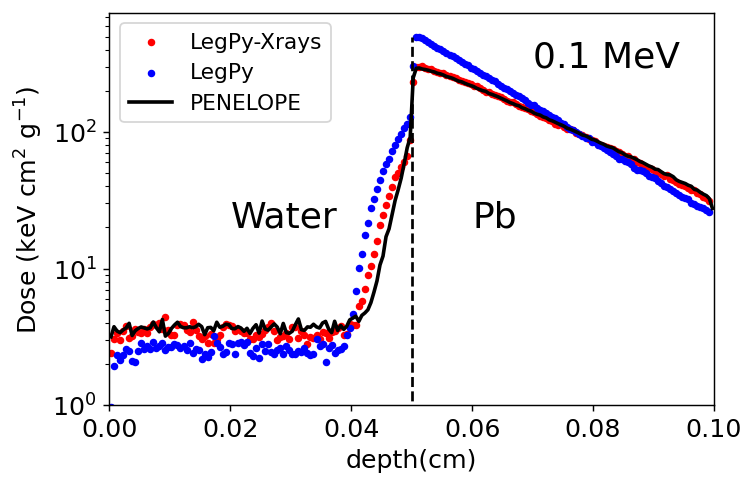}
\caption{
Dose in depth for a beam of 100 keV photons on a water - lead cylinder along the z axis. The dashed line indicates the position of the boundary between the two media. The implementation of the generation of X-rays in LegPy (red points) improves significantly the agreement with PENELOPE (continuous line). See text for details.}
\label{fig:water-pb}
\end{figure}

We studied the effect of the X-ray fluorescence in the shape of the dose in depth function around a water-lead interface for a photon beam of 100 keV. To this end, we simulated a pencil beam crossing a cylinder of 1 mm height (depth) along its axis $z$. The lower half of the cylinder ($0 < z < 0.5$ mm) is water and the upper half ($0.5 < z < 1.0$ mm) is lead. In figure \ref{fig:water-pb} we show the results of $D(z)$. It can be appreciated that the agreement of LegPy with PENELOPE is significantly better at both sides of the interface when our X-ray fluorescence model is included in LegPy.   

\subsection{Absorption spectra in scintillators}
\label{scint}

The knowledge of the distribution of absorbed energy in a scintillator is necessary for a correct interpretation of X-ray and gamma-ray spectra. As is well known, photons that undergo photoelectric absorption contribute to the spectrum photopeak, while those escaping the scintillator after one (or more) Compton scattering contribute to the Compton profile. 
Due to fluorescence, photoelectric absorption produces monoenergetic X-rays that, if escape the scintillator, might give rise to secondary peaks.

\begin{figure}[h]
\centering
\includegraphics[width=\linewidth]{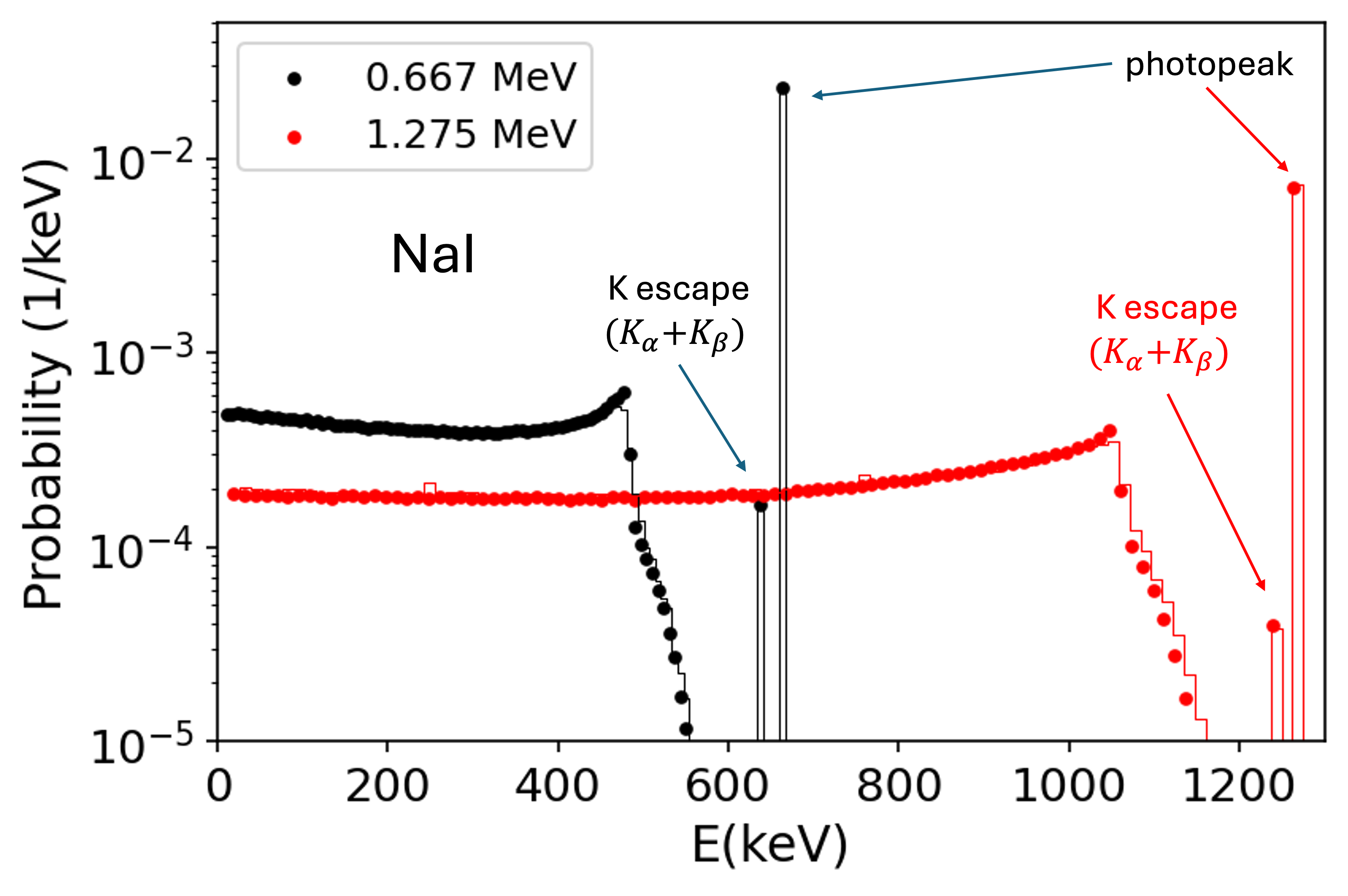}
\caption{Spectra of absorbed energy for 0.667 and 1.275 MeV photons in NaI. Results from the upgraded version of LegPy (points) are compared with those from PENELOPE (continuous lines). The escape peaks due to Iodine X-rays are clearly visible. See text for details.}
\label{fig:NaI}
\end{figure}

In \cite{legpy}, we found that the spectrum of absorbed energy in NaI obtained with LegPy lacked an escape peak that was well observed in the spectrum obtained with PENELOPE. We have repeated the simulation using the upgraded version of LegPy. In figure \ref{fig:NaI}, it is shown the spectrum for isotropic beams of 0.667 and 1.275 MeV emulating radioactive point sources of $^{137}$Cs and $^{22}$Na at a distance of 2 cm from a NaI cylinder of both diameter and height equal to one mean free path (4.34 cm for $^{137}$Cs and 5.32 cm for $^{22}$Na). For the case of 0.667 keV, a secondary peak is observed at 0.64 keV due to the escape of $K_{\alpha}$ and $K_{\beta}$ photons of Iodine with energies of 29 and 32 keV, respectively. The LegPy results are in good agreement with those obtained with PENELOPE. For 1.275 MeV photons, the escape peak is at 1.24 MeV. The agreement is also very good. The only deviation with respect to PENELOPE is the absence of the weak single and double escape peaks at 0.76 and 0.25 MeV due to the escaping of one or two 0.511 MeV photons from positron annihilation, which LegPy ignores so far.

\begin{figure}[h]
\centering
\includegraphics[width=\linewidth]{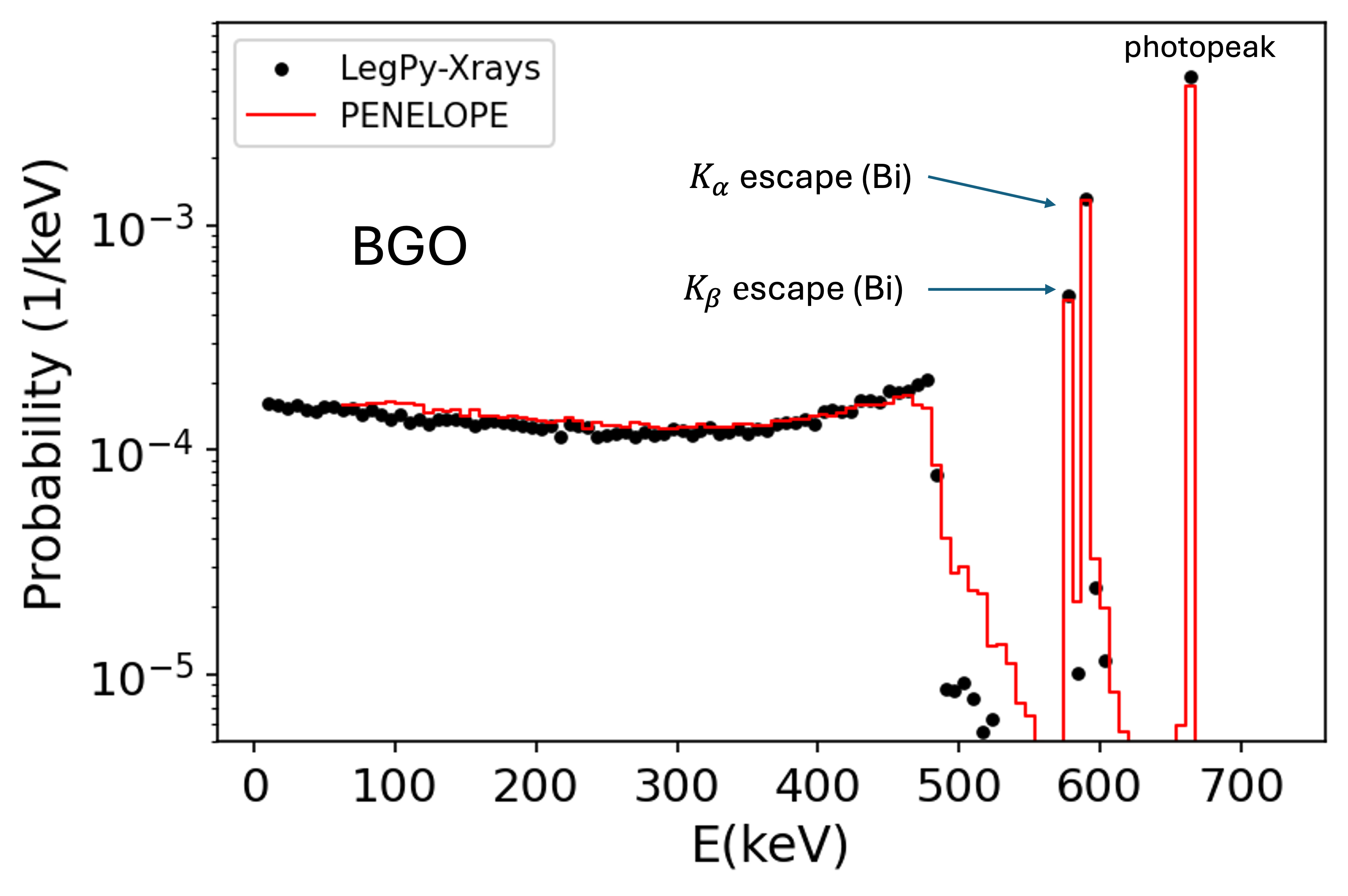}
\caption{Spectrum of absorbed energy for 0.667 MeV photons in BGO. Results from the updated version of LegPy (points) are compared with those from PENELOPE (continuous line). The escape peaks are clearly visible. See text for details.}
\label{fig:BGO}
\end{figure}

A similar comparison has been done for Bismuth Germanium Oxide (BGO) with a $^{137}$Cs source (0.667 MeV). In this case the simulated scintillator has also a size of one mfp (0.107 cm). The results are shown in figure \ref{fig:BGO}. In this case, the X-ray fluorescence from Bismuth is very relevant. The energy of the $K_{\alpha}$ and $K_{\beta}$ X-rays are 76 and 88 keV, respectively, leading to two escape peaks at 579 and 591 keV of significant intensity. The LegPy results are in very good agreement with those of PENELOPE regarding the shape of the Compton profile, the intensity of the photopeak and the intensities of the two escape peaks.    

\section{Conclusions}
\label{concl}
The LegPy package has been upgraded by including the production of X-rays due to atomic de-excitation subsequent to photoelectric absorption. A simple model has been applied in which both the probability of photoelectric absorption and the fluorescence yield are tabulated as a function of the atomic number \cite{Attix}. Only $K_{\alpha}$ and $K_{\beta}$ emissions have been considered. For composed materials, it has been assumed that the relative probability of photoelectric interaction in each atom is proportional to $Z^{3.5}$.

This simple model has been sufficient to fix severe deviations of LegPy results that takes place when the photon energy is slightly larger that of the $K$ shell of some of the atomic components of the medium. In particular, LegPy results for dose in depth for 100 keV photons on lead are now in good agreement with PENELOPE, with deviations less than 10\% at 5 mfp, while neglecting X-rays overestimates the dose up to a factor of about 5. Even in lighter media like bone, Ca X-rays play a non-negligible role as shown in figure \ref{fig:bo_4.71}. We have also found a significant improvement in the performance of LegPy for the calculation of the dose in a water-lead interface at 100 keV (figure \ref{fig:water-pb}).

Only the K shell has been considered in our model, since the effects of X-rays from other shells are expected to be only relevant in few specific cases of limited practical interest. For instance, the dose at a depth of 5 mfp in lead for photons with energy of 20 keV, slightly over the L-shell binding energy (15.9 keV), is underestimated by up to 30\% in 5 mfp if ignoring X-rays. This is still a minor limitation of LegPy in the current implementation. 

Lastly, our model allows LegPy to get accurate results on the spectra of absorbed energy in scintillators, like NaI or BGO, even at photon energies for which X-ray production is very efficient.

\section*{Acknoledgements}
We thank Francesc Salvat for providing help and guidance for the use of PENELOPE program. We gratefully acknowledge financial support from the Spanish Research State Agency (AEI) through the grant PID2019-104114RB-C32. V. Moya also acknowledges the research grant CT19/23-INVM-109 funded by NextGenerationEU and the contract from Comunidad de Madrid (PIPF-2023 /TEC-29694).

\end{document}